# Dipole radio-frequency power from laser plasmas with no dipole moment

F. S. Felber[a]

*Physics Division, Starmark, Incorporated, P. O. Box 270710, San Diego, California 92198*

The radio-frequency power radiated from laser-target plasmas in a vacuum can be orders of magnitude greater than expected from such sources that have a negligible electric dipole moment. A model combining the Tidman-Stamper circuit model of a laser-target plasma with the theory of radiation from currents immersed in plasmas, however, predicts scaling of electric-dipole power radiated from laser plasmas in agreement with experiments.

This paper presents a simple model of electron currents immersed in laser plasmas that estimates the power, angular distribution, and spectrum of radio-frequency (rf) radiation from laser plasmas produced on solid surfaces in vacuum. The estimates of rf power are compared with experimental data from $CO_2$-laser interactions with copper targets at incident intensities from about 60 MW/cm$^2$ to 40 GW/cm$^2$.

The model agrees with the limited observations of rf-power scaling with laser power, with the radial and angular dependence of the near-zone fields, and with the correspondence of rf radiation with the formation of a critical-density surface. The model should be valid over a range of laser wavelengths from the ultraviolet to the long-wavelength infrared. The model assumes quasi-steady-state, one-dimensional ablation, which is valid as long as the laser pulses are not too short and the laser spots are not too small to establish nearly steady ablation across a nearly planar critical-density surface. The model should also be valid over a range of incident laser intensities that produce plasma electron temperatures of a few eV, just above the plasma ignition threshold, to 1 keV or higher. At higher laser intensities, double layers and hot electrons may affect rf production through their dipole moments.

Although laser plasmas have a negligible electric dipole moment at these low intensities, the theory of currents immersed in plasmas suggests that the rf radiation from the current in the diffuse plasma surrounding the dense plasma plume is not cancelled by radiation from the equal and opposite current in the plume. If the rf radiation from such laser plasmas can be understood from a model such as this, then the radiation might serve as a diagnostic tool for laser plasmas. For example, the rf radiation can be used as a sensitive indicator for the threshold of formation of critical-density surfaces.

The first observations of voltages on isolated laser targets [1] and of rf emissions from laser plasmas [2] were tentatively attributed to charge-separation (capacitive) electric fields. At about the same time, nonthermal microwaves observed from a plasma focus were attributed to the Buneman instability in the pinched plasma [3]. Microwave emissions from a plasma point and a plasma focus were investigated to study localized high-temperature plasma formations in high-current pinched discharges [4,5]. Observations of rf emissions during the sublimation of metal targets by a 1 ms, low-intensity laser were attributed to the dipole moment of a double layer at the front of the turbulent expanding metal vapor [6]. At much higher laser intensities of $4 \times 10^{12}$ to $10^{15}$ W/cm$^2$, the first direct indications of double layers in laser plasmas were inferred from measurements of plasma potentials and charged particle currents [7].

Measurements of rf radiation and electron emission current densities from aluminum laser targets were made in the range of laser intensities of interest for our model with $10^{11}$ W/cm$^2$, 10 ps pulses to scale the rf power with the peak intensity of each of the pulses in a mode-locked-laser pulse train [8]. Radio-frequency electric currents were produced on metal targets in air at the 600-MHz modulation frequency of a biharmonic laser at $10^9$ to $10^{10}$ W/cm$^2$ [9]. Analysis of radiation at harmonics of electron waves in plasmas suggests that momentum exchange between electrons and ions can produce dipole power that is comparable to, or somewhat greater than, the quadrupole power, but not by many orders of magnitude [10].

The model of this letter is based on a simple circuit model of a laser plasma [11,12]. This Tidman-Stamper circuit model estimates the voltage drop $V$ along the length of a dense laser-plasma plume to be equal to $T/e$, where $T$ is the plasma electron temperature in energy units, and $e$ is the electron charge. This letter combines the Tidman-Stamper model with the theory of an antenna current immersed in a plasma to predict rf radiation from single laser pulses incident on solid targets. The combined model treats the overdense laser plasma as a short monopole antenna on a ground plane.

As in Ref. 11, the plasma plume is treated as a short cylinder of radius $a$ and length $d$ with a voltage drop $V = T/e$ between the target surface at $z = 0$ and the critical-density surface at $z = d$, as shown in Fig. 1. And as in Ref. 11, the current density $I/\pi a^2$ is assumed to be radially uniform within the cylinder. The circuit equation of Ref. 11 then leads to a total current

$$I \approx V\left(R + \frac{2L}{\tau}\right)^{-1} \approx \frac{T}{e}\left(\frac{d}{\pi a^2 \sigma} + \frac{2d}{\tau c^2}\right)^{-1} . \qquad (1)$$

In this circuit equation, the resistance $R$, in terms of conductivity $\sigma$, is $d/\pi a^2 \sigma$. The contributions of the inductance $L$ to the voltage drop are considered to arise about equally from fluid expansion and from rate of change of current during the laser pulse, when both occur about on the timescale of the laser pulse duration $\tau$.

The combined model determines the plasma electron temperature by a crude partitioning of the incident laser power $P_L = \pi a^2 \Phi$, where $\Phi$ is laser flux. The fraction $f$ of incident laser power that contributes to heating electrons is assumed to be a constant about equal to 0.2. The rest of the laser power is partitioned, without consequence to the model, among laser reflection, hydrodynamic expansion, dissociation, ionization, radiation, etc.

---

[a] Electronic mail: starmark@san.rr.com

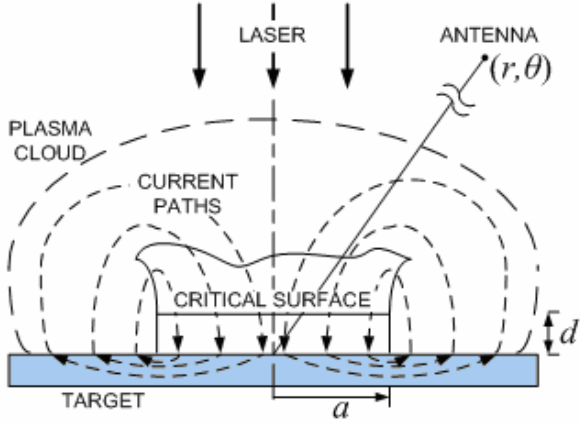

FIG. 1. Illustration of model geometry showing dense plasma plume rising from target surface and poloidal currents immersed in plasma (adapted from Ref. 14).

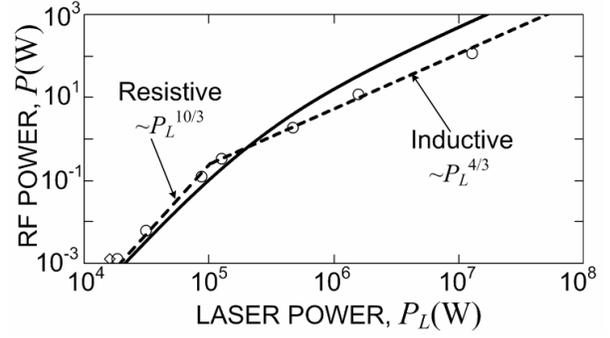

FIG. 2. Peak rf power (watts) from solid copper target vs peak incident $CO_2$-laser power (watts). Data points (circles) and plasma formation threshold (diamond) are from Ref. 19. Model prediction (solid curve) and rf power scalings in resistive- and inductive-voltage regimes (dashed lines) are from Eq. (4).

Electron thermal energy is transported from the overdense region through the critical surface, at which the plasma frequency equals the laser frequency, by plasma at mass density $\rho_c$ and at speed $v_c$. For quasi-steady-state ablation, this outward flux of electron thermal energy is balanced by the fraction $f$ of incident laser energy flux that flows inward from the critical surface, where the laser energy is absorbed, and heats electrons. Since the critical surface is a sonic point, the electron temperature at the critical surface is [13]

$$T = m_i v_c^2 \approx m_i (2f\Phi/3\rho_c)^{2/3}, \qquad (2)$$

where $m_i$ is the ion mass.

As shown in Fig. 1, this azimuthally symmetric current flows down the dense plasma plume and returns through the diffuse plasma surrounding the dense plume [12,14]. The current distribution resembles a poloidal current on a "fat" torus (for which the minor radius is about equal to the major radius). Ordinarily, this current distribution would radiate power of the same order as a magnetic dipole or electric quadrupole [15]. Then the rf power would be expected to be smaller than the power from an electric dipole by a factor of order $(ka)^2 \ll 1$, where $k$ is the rf wavenumber. But the measured rf power is orders of magnitude greater than that from a magnetic dipole or electric quadrupole. Electric-dipole power seeming to be radiated from a laser plasma that has negligible dipole moment may be explained by the following mechanism.

*A system of n-poles having zero total n-pole moment will generally produce n-pole radiation if the fields of the individual n-poles are modified differently in the source region.*

An antenna in a plasma produces an electroacoustic wave, as well as an electromagnetic wave. The radiation resistance $R_{rad} = R_p + R_e$ of a short, thin cylindrical antenna in a plasma comprises an electroacoustic component $R_p$ and an electromagnetic component $R_e \approx R_0 (1 - \omega_p^2/\omega^2)^{1/2}$ [16]. Here, $\omega_p$ is the electron plasma (angular) frequency, and $R_0 \approx 4(kd)^2/3c \; [= 40(kd)^2 \text{ ohms}]$ is the free-space ($\omega_p = 0$) radiation resistance of a very short monopole antenna on a reflecting plane with uniform current along its length $d$ [17]. As $\omega_p$ approaches $\omega$, the electroacoustic radiation resistance, which scales as $R_p \sim (\omega_p/\omega)^2/(1-\omega_p^2/\omega^2)$, becomes very large, from which one concludes that nearly all power goes into the electroacoustic wave under such conditions [16,18].

The theory suggests, therefore, that electromagnetic radiation does not escape from a short thin antenna immersed in an overdense plasma [16]. (Here, "overdense" means the plasma frequency $\omega_p$ is greater than the rf angular frequency $\omega$, not the laser frequency.) But the return current, which flows in the diffuse plasma surrounding the plume, does radiate with a dipole moment rate $Id$. The net effect is that the laser plasma radiates as an electric dipole, instead of a quadrupole. Radio-frequency radiation can still escape from the diffuse plasma carrying the return current even if this diffuse plasma is overdense with respect to the rf frequency, as long as the characteristic dimensions of the diffuse plasma are less than the plasma skin depth.

If the angular frequency of the half-cycle voltage pulse is taken to be $\omega_0 = \pi/2\tau$, then the electric dipole moment of the return current is

$$p = Id/2\omega_0 \; . \qquad (3)$$

A laser plasma produced on a timescale $\tau$ should radiate broadband rf at frequencies around $1/4\tau$ as a short, thin monopole antenna on a ground reflecting plane. The rf power radiated above the target, with the angular distribution of an electric dipole, is

$$P = \frac{\omega_0^4}{6c^3}|p|^2 \approx \frac{\pi^2 m_i^2 c}{384 e^2}\left(\frac{2f\Phi}{3\rho_c}\right)^{4/3}\left(1 + \frac{\tau c^2}{2\pi a^2 \sigma}\right)^{-2} \; . \qquad (4)$$

Using Eqs. (1) through (4), and noting that the critical plasma mass density $\rho_c$ scales with laser wavelength $\lambda$ as $\lambda^{-2}$ and Spitzer conductivity scales as $T^{3/2}$, we find the scaling of rf power shown in Fig. 2. Figure 2 compares the absolute model predictions of rf power with $CO_2$-laser experiments [19]. The experiments used 3-ns to 4-ns-wide pulselets from a mode-locked $CO_2$ transversely-excited-atmospheric-pressure laser focused through various optical attenuators to a 200-μm-diameter spot on solid copper targets immersed in a vacuum of about $10^{-4}$ Torr [19].

The model predicts that the frequency of fundamental-mode rf emissions for these experiments should have a peak in the low VHF about $1/4\tau \approx 60 - 80$ MHz. But the antennas measured only frequencies $\geq 250$ MHz in the high VHF and above [19]. If the antennas were measuring fundamental-mode rf power below the roll-off in their frequency response or were measuring third or higher harmonics, then the data in [19] should underestimate the actual rf power.

Although error bars are not available, the experiments of Ref. 19 seem to support the model scaling of rf power with peak incident laser power, and seem to show the change in scaling from resistive-voltage to inductive-voltage laser plasmas, as seen by the dashed lines in Fig. 2. The experiments also confirmed that rf emissions were observed only when a critical density surface was formed [19].

If this model correctly explains dipole radiation from a source having a negligible dipole moment, then the door is open for applications beyond mere diagnostic tools. For example, one might imagine compensating for atmospheric turbulence or steering a beam by nonlinear optical control of the phase of a beam over an optical surface through individual surface elements that have no moving parts.

The author is grateful to Neal Carron and John Stamper for helpful discussions and communications.